\documentclass[twocolumn,preprintnumbers,amsmath,amssymb]{revtex4}

\usepackage{graphicx}  																																																									
\begin{document}
\title{Modulation characteristics of uncooled graphene photodetectors
}
\author{V. Ryzhii$^{1,2}$,   M. Ryzhii$^3$, T.~Otsuji$^1$,
V. Leiman$^4$, 
V. Mitin$^5$, 
and M. S. Shur$^6$}
\address{
$^1$Research Institute of Electrical Communication, Tohoku University, Sendai 980-8577, Japan\\
$^2$ Institute of Ultra High Frequency Semiconductor Electronics of RAS,
Moscow 117105, Russia\\
$^3$Department of Computer Science and Engineering, University of Aizu, Aizu-Wakamatsu 965-8580, 
Japan\\
$^4$Center for Photonics and Two-Dimensional Materials, Moscow Institute of Physics
and Technology, Dolgoprudny 141700, Russia\\ 
 $^5$Department of Electrical Engineering, University at Buffalo, SUNY, Buffalo, New York, 1460-192 USA\\
$^6$Department of Electrical, Computer, and Systems Engineering, Rensselaer Polytechnic Institute, Troy, New York 12180, USA
}
 \begin{abstract} 
%\noindent
%{\bf Keywords:}  graphene,   perforated graphene 
 %photodetector.\\
 We analyze the modulation characteristics of the uncooled  terahertz (THz) and infrared (IR) detectors using the variation of the density and effective temperature of 
the two-dimensional electron-hole plasma  
in uniform  graphene  layers (GLs) and perforated  graphene  layers (PGLs) 
due to the absorption of THz and IR radiation. 
The performance of the   photodetectors (both the GL-photoresistor and the PGL-based barrier photodiodes) are compared.
Their characteristics are also compared with the GL reverse-biased photodiodes. 
 The obtained results allow
to evaluate the ultimate modulation frequencies of these photodetectors 
and can be used for their optimization. 
\end{abstract} 

\maketitle

\newpage

\section{INTRODUCTION}

The non-equilibrium two-dimensional electron-hole plasma (2DEHP) in graphene layers (GLs)  from equilibrium~\cite{1,2,3,4,5,6,7,8,9} enables
the  operation  of
different GL-based terahertz (THz) and infrared (IR)
 electro-optical  modulators, ultrafast thermal light sources with high carrier effective temperature,   the superluminescent and lasing diodes, and, particularly, 
the 
hot/cool  carrier bolometric detectors  
of electromagnetic radiation~\cite{10,11,12,13,14,15,16,17,18,19,20}.
Such detectors use the effect of the conductivity
variation due to the carrier density and effective temperature
variation and the change associated with the radiation absorption. %The zero-gap energy spectra of GLs  The interband incident  photon absorption strongly contributes to the carrier generation, heating, or cooling.
 The speed (in particular, the maximum modulation frequency) of the GL-based photodetectors is one of their important characteristics.
The maximum modulation frequency depends on not only  the carrier energy relaxation and recombination,  but also on the 2DEHP heat capacities. 
Despite a relatively small value of the latter compared to the heat capacity   of the substrate and of  the GL encapsulation  layers in which GLs are encapsulated, it can limit the  modulation speed.

In this paper, we study the modulation characteristics
of the uncooled  THz and IR GL hotodetectors  using the effects of the interband carrier generation/recombination and 
 the carrier heating/cooling (the bolometric mechanism). We account for the essential interdependence of these processes. 
We clarify of the relative roles of the carrier energy accumulation  and the
relaxation and recombination processes in the dynamic (modulation) properties
of the uncooled (operating at room temperature) photodetectors. Our analysis  accounts for 
the correct value of the 2DEHP capacity, which substantially deviates from its classical value.

The developed device model describes the high-speed operation of the
photodetectors with
uniform and perforated GLs.

\begin{figure*}{}
\centering
\includegraphics[width=10.5cm]{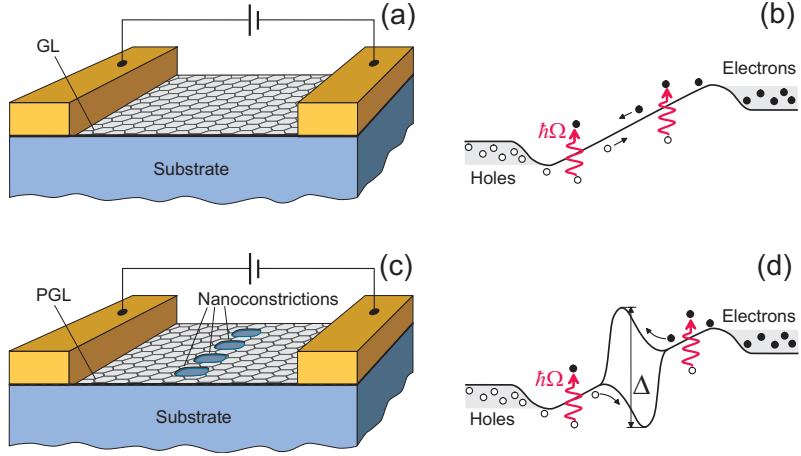}
\caption{Schematic view of the device structures and their band diagrams: (a) and (b) - for GL-photoresistor with uniform GL channel, and (c) and (d) - for
 PGL-photodiode with the barrier-limited carrier transport via the perforated GL region (with the barrier heights in the nanoconstrictions $\simeq \Delta/2$). Vertical wavy arrows indicate the interband absorption of photons with energy $\hbar\Omega$. }
\label{F1}
\end{figure*}

\section{DEVICE MODEL AND GENERAL EQUATIONS}

We consider the photodetectors based on the undoped GLs  supplied by the side Ohmic (p- and n-) contacts. The Fermi levels in the contacts  align with the GL Dirac
points shifted,
by the dc bias voltage $V$. The GL  regions can be either
uniform (in the GL photoresistors) or comprise a perforated  array  in the perforated GL (PGL) barrier photodiodes. 
Figure 1 show the  structures and the band diagrams of the devices under study. 
The normally incident  THz or IR radiation absorbed in the GL 
 results in the extra electron-hole pair generation and affects the  energy of the 2DEHP. 
 The
variation of the carrier density and their effective temperature leads to the variation of the GL  conductivity and the current through the PGL region and modulates the terminal current. 
In the perforation array, the carrier current flows only in the nanoconstrictions~\cite{21,22,23,24} (GL quantum wires or nanoribbons) between the perforations. If the nanoconstrictions are sufficiently narrow, the quantization of the carrier energy spectrum results in the barriers for the carriers coming from the uniform parts of the GL, which limit the terminal current. 
The steady-state bolometric characteristics of the uniform GL
and the GLs with the region patterned into the GL nanowires (nanoribbons) were evaluated previously~\cite{9,12}. 
As it was predicted a long time ago~\cite{21}, the inclusion of the vertical and lateral barrier regions between the photosensitive standard quantum wells ~\cite{25,26,27} and 
between the GLs in 
the  GL-based hot-carrier bolometers with the vertical structure~\cite{28,29}  generally leads to the responsivity enhancement. The barriers between the carrier puddles and the lateral quantum dots in GLs~\cite{13,16} can play a similar role.

 The 2DEHP density and effective temperature are determined by  radiation absorption and the interband (recombination) and intraband relaxation processes.

Due to relatively frequent carrier-carrier collisions in the GLs  at room temperatures, 
 the 2DEHP electron and hole subsystems can be described by quasi-Fermi energy distribution
functions $f_e(\varepsilon)$ and $f_h(\varepsilon)$ with common effective temperature $T$ (in energy units)
:
$f_e(\varepsilon)=\displaystyle
\biggl[1 + \exp\biggl(\frac{\varepsilon - \mu_e}{T}\biggr) \biggr]^{-1}$, $f_h(\varepsilon)= \displaystyle \biggl[1 + \exp\biggl(\frac{\varepsilon - \mu_h}{T}\biggr) \biggr]^{-1}$. Here $\varepsilon \geq 0$ is the carrier kinetic energy and $\mu_e$ and $\mu_h$
are the electron and hole quasi-Fermi energies, respectively. 
Due to a symmetry of the electron and hole subsystems in the situations under consideration, $\mu_e = \mu_h =\mu$. In the equilibrium (without an external  irradiation),
$\mu = 0$. Generally, the absorption of the incident radiation can lead to $\mu \neq 0$, although, except for the cases of a strong irradiation, $|\mu| \ll T$.

Limiting our consideration to the room temperatures, we assume that the main mechanisms of the density and energy relaxation are associated with the GL  optical phonons~\cite{30,31,32}. 
%Due to the features of the operation conditions (not to high carrier densities %temperatures, we neglect the contribution of the Auger processes~\cite{26} (see, %however, Ref.~\cite{27,28}).

The pertinent rate equations, which govern the density and energy balances in the 2DEHP, are given by

\begin{eqnarray}\label{eq1}
\frac{d \Sigma}{d t} = G - R^{inter} - R^{A}, 
\end{eqnarray}

\begin{eqnarray}\label{eq2}
\frac{d {\cal E}}{d t} = \hbar\Omega\,G - \hbar\omega_0(R^{inter} + R^{intra}). 
\end{eqnarray}
Here $\Sigma$ and ${\cal E}$ are the net carrier density and the carrier energy
 per unit square, respectively,  
 in the entire GL or its parts separated by the constriction area,
 $G$, $ R^{inter}$, $R^{intra}$, and  $R^A$ are the pertinent rates of the electron-hole pair generation by the incoming radiation and  the interband and intraband transitions with the generation and absorption of optical phonons, 
 and the Auger recombination-generation processes~\cite{9,33,34,35}, respectively,
$\hbar\Omega$ and $\hbar\omega_0$ are the photon and optical phonon energies ($\hbar\omega_0 \simeq 200$~meV), and $\hbar$ is the Planck constant. 
The right-hand side of Eq.~(2) reflects the fact that each absorbed photon
increases the 2DEHP energy by a quantity $\hbar\Omega$, while the absorption and emission of optical phonons change this energy by $\pm\hbar\omega_0$.
The Auger processes affect the carrier density, however, due to the zero energy gap, they
do not vary  the 2DEHP energy. 
The explicit expressions for the relaxation terms $R^{inter}$, $R^{intra}$, and  $R^{A}$ are given (exemplified) in the Appendix A.

  The dispersion relations for electrons (upper sign) and holes (lower sign)
 in the GLs are presented as
 
\begin{eqnarray}\label{eq3}
\varepsilon^{\pm} = \pm v_Wp, 
\end{eqnarray}
where $v_W \simeq 10^8$~cm/s is the characteristic carrier velocity in GL , $p = |p|$ is the carrier momentum. Hence, for the rate of the electron-hole pairs photogeneration we arrive 
 at the following formulas:

\begin{eqnarray}\label{eq4}
 G= \beta\, \tanh\biggl(\frac{\hbar\Omega - 2\mu}{4T}\biggr)\,I_{\Omega}(t).
\end{eqnarray}
Here   
$\beta = \pi\,e^2/c\hbar\sqrt{\kappa} = (\pi/137\sqrt{\kappa})$,  $e$ and $c$ are the electron charge and light speed, respectively, $\kappa$ is the dielectric constant of the layers surrounding the GL, $T$ is the 2DEHP temperature. 
The normally incident radiation flux, $I_{\Omega}(t)$,
comprises the steady-state component $I_{\Omega}$ and the modulation signal component  $I_{\Omega}^{\omega}\exp(-i\omega t) $: 
$$I_{\Omega}(t) = I_{\Omega} + I_{\Omega}^{\omega}\exp(-i\omega t),
$$ 
where $I_{\Omega}^{\omega}$ and 
 $\omega$ are the amplitude  and the frequency of the modulated signal.

The absorption the THz and IR photons  in GLs  is due to the interband (mainly direct optical  transitions) and to the intraband (indirect transitions). At the photon frequencies $\Omega \gg \tau^{-1}$,
where $\tau$ is the carrier momentum relaxation time, the latter mechanism, which corresponds to the carrier Drude absorption, is relatively weak and, for simplicity, it is disregarded in Eq.~(3) (although this mechanism was included into our model previously~\cite{9}).
The carrier densities  and their energy densities as functions of the effective temperature and the quasi-Fermi energy are  given in  the Appendix B, Eq.~(B1).

The GL  conductivity is determined by the carrier densities, i.e., by the quasi-Fermi energy $\mu$, and by their effective temperature $T$.
Assuming that the GL  dc and low signal frequency conductivities, $\sigma_{GL}$  are determined by the short-range scattering (on the defects and acoustic phonons), 
so that  $\tau = \tau_0 (T_0/pv_W) \propto 1/p$ (for GLs)~\cite{3,9}, where 
 $\tau_0$ is the characteristic short-range scattering time~\cite{36}, one can obtain the following  formula:

\begin{eqnarray}\label{eq5}
\sigma_{GL} = \frac{2\sigma^{dark}}{1 + e^{-\mu/T}}.
\end{eqnarray}
Here
$\sigma^{dark} = (e^2T_0\tau_0/\pi\hbar^2) = \sigma_{00}$ is the intrinsic conductivity
in the dark condition: $T = T_0$ and $\mu = 0$. 
At  relatively weak irradiation,  $|\mu| \ll T_0,  T$, and Eq.~(5) yields

\begin{eqnarray}\label{eq6}
\sigma_{GL}\simeq \sigma^{dark}\biggl(1 + \frac{\mu}{2T}\biggr).
\end{eqnarray}
According to Eq.~(6), the terminal photocurrent across the uniform GL, $\Delta J_{GL} = J_{GL}- J_{GL}^{dark}$,  between the side contacts in the GL  photoresistors can be presented
as
\begin{eqnarray}\label{eq7}
\Delta J_{GL}
\simeq \sigma^{dark}\frac{VH}{4L}\frac{\mu}{T}, 
\end{eqnarray}
 so that its value normalized by the dark currents is given by

\begin{eqnarray}\label{eq8}
\frac{\Delta J_{GL}}{J_{GL}^{dark}} \simeq \frac{\mu}{2T_0}.
\end{eqnarray}
Here  $2L$ is the spacing between the contacts and $H$ is the GL width in the direction
perpendicular to the current.

The nanoconstriction~\cite{21,22} constitutes a 1D channel for electrons and holes (quantum wires or nanoribbons~\cite{37,38,39}).
Considering that the energy gap in the nanoconstriction as a function of the coordinate $z$
along the carrier propagation direction is determined by 
the nanoconstriction length and width, $2l$ and $d$, respectively, and  
the voltage drop between the nanoconstriction edges $V_{pn} \lesssim V$: $\Phi(z) = \Phi^{max}(1 - z^2/l^2)$, can be presented as
$\Phi^{max}\simeq (\Delta \mp eV_{pn})/2$ (if $eV_{np} \ll \Delta$) with
$\Delta = 2\pi\hbar\,v_W/d$.
For an effective control of the net current the height of the barrier (i.e., the
energy gap $\Delta$) should be sufficiently large to provide relatively small voltage drop across the uniform regions in  the GL and GBL. In this case, $V_{pn} \simeq V$,
where $V$is the bias voltage between the side contacts.
Accounting for this, the ratio of the net thermionic photocurrents  currents, $\Delta J_{PGL}$ and $\Delta J_{PGB}$, in the photodiodes based on the perforated GLs (PGLs) over the barriers in  the nanoconstriction~\cite{22},  can be presented as

\begin{eqnarray}\label{eq9}
\frac{\Delta J_{PGL}}{J_{PGL}^{dark}}  \simeq 
\biggl(\frac{\Delta}{2T_0}\biggr)
\biggl[\frac{\mu}{T_0} + \frac{(T-T_0)}{2T_0}\biggr].
\end{eqnarray}

One can see from the comparison of Eqs.~(8) and (9) that the 
relative variations of the net currents under the irradiation
 in the case of the perforated GLs  comprise the factor
$\Delta/2T_0 $, which can be large for  sufficiently narrow nanoconstrictions.

\section{STEADY-STATE CARRIER QUASI-FERMI ENERGY, EFFECTIVE TEMPERATURE, and DC PHOTOCURRENT}

At  $|\mu| \ll T \ll \hbar\omega_0$,
Eqs.~(1) - (3) with  Eqs.~(A1) and (B1)  result in  the following
values of the steady-state components of the quasi-Fermi energies  $ \mu$
and the effective temperature $T$:

\begin{eqnarray}\label{eq10}
\frac{ \mu}{T_0} = \frac{\beta\biggl(1 +a - \displaystyle\frac{\Omega}{\omega_0}\biggr)}{2(a +b+ab)} \tanh\biggl(\frac{\hbar\Omega}{4T_0}\biggr)\frac{ I_{\Omega}}{I}, 
\end{eqnarray}

\begin{eqnarray}\label{eq11}
\frac{ T - T_0}{T_0} \simeq \frac{\beta\biggl(\displaystyle\frac{\Omega}{\omega_0}-1\biggr)\biggl(\frac{T_0}{\hbar\omega_0} \biggr)}{(a + b+ab)}
\tanh\biggl(\frac{\hbar\Omega}{4T_0}\biggr)\frac{ I_{\Omega}}{I}.
\end{eqnarray}
Here $a = \displaystyle\biggl(\frac{\pi\,T_0}{\hbar\omega_0}\biggr)^2\biggl(1 + \frac{2.19T_0}{\hbar\omega_0}\biggr)$,
$b = t_0/t_A$, $t_0 = \Sigma_0/I$, $\Sigma_0 = (\pi/3)(T_0/\hbar\,v_W)^2$ is the 2DEHP density at $T = T_0$, $t_A$ is the characteristic time of the Auger carrier generation,
 $I$ is the rate of the electron-hole pair generation in GLs due to the absorption of equilibrium optical phonons $I \simeq (1-2)\times 10^{21}$~cm$^{-2}$s$^{-1}$~\cite{21}. At $T_0 = 25$~meV, $a \simeq 0.196$ and $t_0 \simeq 84 - 168$~ps.

Equations~(10) and (11) show that depending on the ratio $\Omega/\omega_0$, the quasi-Fermi energies and the deviation of the effective temperatures from the lattice temperature  can be both positive and negative. The radiation frequencies, at which the quasi-Fermi energies change
 their signs are the same ($\Omega = \omega_0$), 
while the change of the signs of   $(T_{GL} - T_0)$ and $( T_{PGL} - T_0)$ occurs at
  somewhat different $\Omega$.   

Omitting the term $b$ due to its smallness (see Appendix A), the dc photocurrents in the GL photoresistor and PGL photodiode can be presented as

\begin{eqnarray}\label{eq12}
 \frac{\Delta J_{GL}}{J_{GL}^{dark}} \simeq 
\frac{\beta}{4a}\biggl(1 +a - \frac{\Omega}{\omega_0}\biggr)
\tanh\biggl(\frac{\hbar\Omega}{4T_0}\biggr)\frac{I_{\Omega}}{I},
\end{eqnarray}

\begin{eqnarray}\label{eq13}
 \frac{\Delta J_{PGL}}{J_{PGL}^{dark}} = \frac{\beta}{2a}\biggl(\frac{\Delta}{2T_0}\biggr) 
 \,\biggl[1 +a - \frac{\Omega}{\omega_0}\nonumber\\
+ \biggl(\frac{T_0}{\hbar\omega_0}\biggr)\,\biggl(\frac{\Omega}{\omega_0} - 1 \biggr)\biggr]
\tanh\biggl(\frac{\hbar\Omega}{4T_0}\biggr)\frac{I_{\Omega}}{I}.
\end{eqnarray}
Equation~(12) coincides with the pertinent formula obtained previously~\cite{9} for the case of the short-range carrier scattering. The expression for the GL photoconductivity given by Eq.~(12) does not
explicitly comprises the term associated with the variation of the effective temperature, i.e., the bolometric term. However, the variation of the quasi-Fermi energy and, particularly, the possible change of this quantity sign at a certain ratio $\Omega/\omega_0$, is related to the 2DEHP heating (or cooling). This implies that the bolometric effect in the GL photoresistors plays a role, although implicitly.
In contrast to the GL photoconductivity, the expression for the PGL photoconductivity explicitly comprises the term associated with the bolometric effect. The bolometric contribution is proportional to a small parameter $T_0/\hbar\omega_0 = 0.125$.
The  presence of the energy barrier in the constrictions area, the value of the photocurrent in the PGL photodiode in comparison with the GL photoresistor is proportional to a large factor 
$\Delta/2T_0$.

\begin{figure}[t]
\centering
\includegraphics[width=8.0cm]{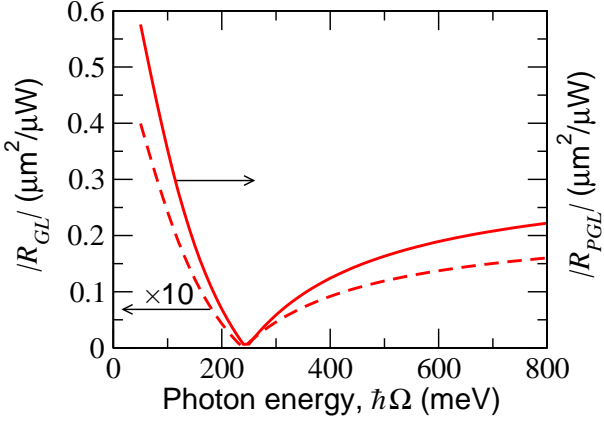}
\caption{ Normalized dc responsivity of a GL photoresistor $|R_{GL}|$ (dashed line)
and  of a PGL photodiode  $|R_{PGL}|$ (solid line) versus photon energy $\hbar\Omega$.}
\label{F2}
\end{figure}
As seen from Eqs.~(12) and (13), the  dc photoconductivity changes the sign at a certain ratio $\Omega/\omega_0$.  
This is because at a large $\Omega$, the radiation absorption
leads to the 2DEHP heating that results in a lowering of the electron quasi-Fermi level (it goes below the Dirac point to the valence band) and in a elevating of the hole quasi- Fermi level (which appears in the conduction band), while at $\Omega/\omega_0 < 1$, the radiation absorption can give rise to the 2DEHP cooling and, hence, to $ \mu < 0$.

Figure~2 shows the spectral characteristics of the GL photoresistor and PGL
photodiode normalized dc responsivities, 
$|R_{GL}|= |\Delta J_{GL}/J_{GL}^{dark}\hbar\Omega\,I|$ and $|R_{PGL}| = |\Delta J_{PGL}/J_{PGL}^{dark}\hbar\Omega\,I|$, calculated using Eqs.~(12) and (13). We
 set $\kappa = 4$, $a = 0.196$, and $\Delta = 400$~meV (the nanoconstrictions with $d 
 \simeq 10$~nm).

The most remarkable feature of the spectral characteristics of $|R_{GL}|$ and
$|R_{PGL}|$ seen in Fig.~2 is that these values turn to zero at a certain ratio 
$\Omega/\omega_0$. This is associated with the equality of the energy received by the 2DEHP from the absorbed radiation and the energy transmitted to the optical phonons
when $\Omega/\omega_0 \simeq 1$.  As follows from Eqs.~(10) and (11), at the latter
relation, $\mu$ and $T-T_0$ become equal to zero, that leads to the zero values of the photocurrent.

\section{MODULATED QUASI-FERMI ENERGY, EFFECTIVE TEMPERATURE, AND AC PHOTOCURRENT}

When the incident radiation comprises the ac modulation component 
$I_{\Omega}^{\omega}(t)$, we arrive at the following equations
for the ac components of the quasi-Fermi energy and effective temperature, 
$\mu^{\omega}$ and  $ T^{\omega}$: 

\begin{eqnarray}\label{eq14}
 2(1 -i\omega t^{(1)}) \frac{ \mu^{\omega}}{T_0} +(1 - i\omega t^{(2)})\biggl(\frac{\hbar\omega_0}{T_0}\biggr)\frac{T^{\omega}}{T_0}\nonumber\\ = \beta\displaystyle\tanh\biggl(\frac{\hbar\Omega}{4T_0}\biggr) \frac{ I_{\Omega}^{\omega}}{I},
\end{eqnarray}

\begin{eqnarray}\label{eq15}
2(1 -i\omega t^{(2)}) \frac{ \mu^{\omega}}{T_0} +(1+a)(1 - i\omega t^{(3)})\biggl(\frac{\hbar\omega_0}{T_0}\biggr)\frac{ T^{\omega}}{T_0}\nonumber\\  = \beta
\biggl(\frac{\Omega}{\omega_0}\biggr)\displaystyle\tanh\biggl(\frac{\hbar\Omega}{4T_0}\biggr) \frac{ I_{\Omega}^{\omega}}{I},
\end{eqnarray}
which in the limit $\omega \rightarrow 0$  convert into Eqs.~(10) and (11), respectively.
The characteristic times $t^{(1)}$, $t^{(2)}$, and $t^{(3)}$ are
given in the Tables I and II and are being proportional to the time $t_0=  \Sigma_0/I$ related to the electron-hole pair generation rate associated with the optical phonon absorption in GLs. Assuming that $I = (1-2)\times 10^{21}$~cm$^{-2}$s$^{-1}$ at $T_0 = 25$~meV, for $b \ll 1$ (weak Auger processes)
we obtain $t_0\simeq (84 - 168)$~ps.  The carrier heat capacity (per one carrier)  
$c \simeq 6.58$ [see Appendix B, Eq.~B(6)]

The times $t^{(1)}$ and $t^{(3)}$  are  the effective carrier recombination and cooling/heating times, respectively. 
The inclusion of the Auger recombination leads to a decrease in the recombination time
$t^{(1)}$.
One can see that $t^{(3)}$
is proportional to the product of  the heat capacitance per one carrier~$c$
and
the  carrier energy relaxation time due to the interaction with optical phonons  $t_0(T_0/\hbar\omega_0)^2$ (that is in line with the previous calculation of this time
in the 2D systems~\cite{4}).
Despite a relatively short time $t^{(3)}$, it is longer than one could assume
in the case of the classical value $c=1$ (see the Appendix B).  

For the 2DEHP in the GL, the hierarchy of the characteristic times is as follows:
$t^{(1)} > t^{(2)} > t^{(3)}$. 
One can see from Eqs.~(14) and (15) that  $ \mu^{\omega}$
substantially depends on  $ T^{\omega}$. In particular, an increase in $ T^{\omega}$ leads to a decrease in $ \mu^{\omega}$. Equations~(14) and (15) result in

\begin{eqnarray}\label{eq16}
\frac{ \mu^{\omega}}{T_0}
= \frac{\beta}{2}\,\frac{Z_{\Omega}^{\omega}}{(1-i\omega t^{(1)})}\displaystyle\tanh\biggl(\frac{\hbar\Omega}{4T_0}\biggr)\frac{ I_{\Omega}^{\omega}}{I},
\end{eqnarray}

\begin{eqnarray}\label{eq17}
\frac{ T^{\omega}}{T_0}
= \beta\biggl(\frac{T_0}{\hbar\omega_0}\biggr)\,\frac{(1 - Z_{\Omega}^{\omega})}{(1-i\omega t^{(2)})}\displaystyle\tanh\biggl(\frac{\hbar\Omega}{4T_0}\biggr)\frac{ I_{\Omega}^{\omega}}{I},
\end{eqnarray}
where

\begin{eqnarray}\label{eq18}
Z_{\Omega}^{\omega} =\frac{1 + a - 
\displaystyle\biggl(\frac{\Omega}{\omega_0}\biggr)
\frac{(1 - i\omega\,t^{(2)})}{(1 - i\omega\,t^{(3)}}}
{1 + a - \displaystyle\frac{(1 - i\omega\,t^{(2)})^2 }
{(1-i\omega t^{(1)})(1 - i\omega\,t^{(3)})}}.
\end{eqnarray}

\setlength{\tabcolsep}{8pt}
\begin{table}[t!]
\caption{\label{table1} GL, PGL  parameters (definitions)}
\vspace{2 mm}
\centerline{\begin{tabular}{lcccc} \hline\hline  
 & $t^{(1)}/t_0$ & $t^{(2)}/t_0$ & $t^{(3)}/t_0$   \\ \hline       
GL, PGL  & $\displaystyle\frac{6\ln2}{\pi^2(1+b)}$       
&   $\displaystyle\frac{2T_0}{\hbar\omega_0}$          
& $\displaystyle\biggl(\frac{T_0}{\hbar\omega_0}\biggr)^2\frac{c}{(1 +a)}$ \\ \hline\hline
\end{tabular}}
%\end{table*}
%
%
%\begin{table}[b]
\caption{\label{table2} GL, PGL parameters (numerical values)}
\vspace{2 mm}
\centerline{\begin{tabular}{lcccc} \hline\hline
& $t^{(1)}$~(ps)  & $t^{(2)}$~(ps) & $t^{(3)}$~(ps)   \\ \hline
GL, PGL & 35 -- 70       &   21 -- 42         & 7 -- 14 \\ \hline
\hline
\end{tabular}}
\end{table}

\begin{figure}[b]
\centering
\includegraphics[width=8.0cm]{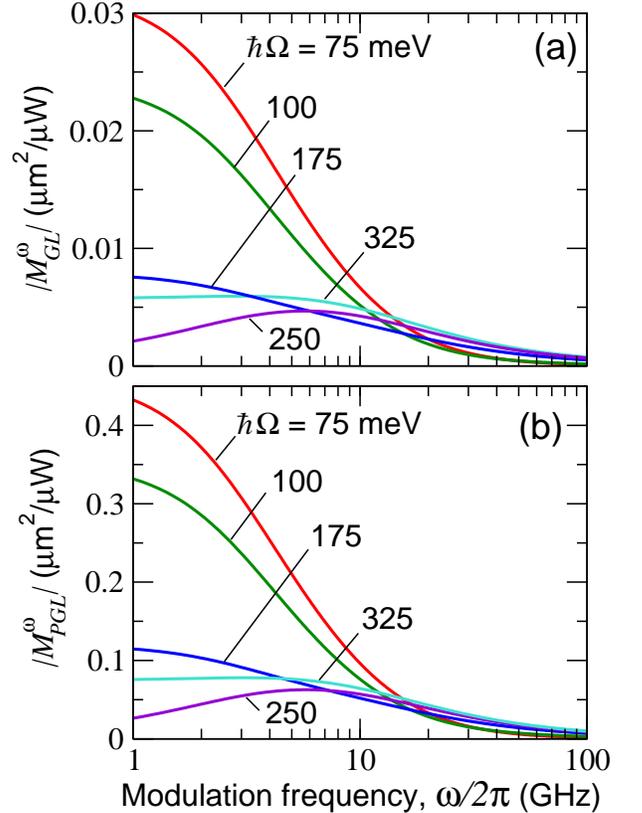}
\caption{ Modulation depth (a) in GL photoresistor $|M_{GL}^{\omega}|$ 
and (b) in PGL photodiode  $|M_{PGL}^{\omega}|$  versus modulation  frequency $\omega/2\pi$ for different 
photon energies $\hbar\Omega$.}
\label{F3}
\end{figure}

Calculating the ac components of the ac photocurrent current
$\Delta J_{GL}^{\omega}$  as in   the previous section, 
 we arrive at the following equation:

\begin{eqnarray}\label{eq19}
 \frac{\Delta J_{GL}^{\omega}}{J_{GL}^{dark}} = \frac{\beta}{4}
\,\frac{Z_{\Omega}^{\omega}}{(1-i\omega t^{(1)})}\displaystyle\tanh\biggl(\frac{\hbar\Omega}{4T_0}\biggr)\frac{\ I_{\Omega}^{\omega}}{I}.
\end{eqnarray}
Analogously, invoking Eq.~(9), for the ac  photocurrent in the PGL photodiodes we obtain

\begin{eqnarray}\label{eq20}
 \frac{\Delta J_{PGL}^{\omega}}{J_{PGL}^{dark}} = \frac{\beta}{2}\biggl(\frac{\Delta_{GL}}{2T_0}\biggr) 
\,\biggl[\frac{Z_{\Omega}^{\omega}}{(1-i\omega t^{(1)})}\nonumber\\
+  \biggl(\frac{T_0}{\hbar\omega_0}\biggr)\,\frac{(1 - Z_{\Omega}^{\omega})}{(1-i\omega t^{(2)})}\biggr]\displaystyle\tanh\biggl(\frac{\hbar\Omega}{4T_0}\biggr)\frac{\ I_{\Omega}^{\omega}}{I}.
\end{eqnarray}

\section{MODULATION DEPTH}

The modulation depths normalized by the dark currents (in units $\mu$m$^2$/$\mu$W) can be defined as

\begin{eqnarray}\label{eq21}
M_{GL}^{\omega} = \biggl(\frac{\Delta J_{GL}^{\omega}}{J_{GL}^{dark}}\biggr)\frac{1}{\hbar\Omega I_{\Omega}^{\omega}}, \,\,
M_{PGL}^{\omega} = \biggl(\frac{J_{PGL}^{\omega}}{J_{PGL}^{dark}}\biggr)\frac{1}{\hbar\Omega I_{\Omega}^{\omega}},
\end{eqnarray}
for the GL photoresistors and the PGL photodiodes, respectively.

Figure~3   shows the absolute value (modulus) of the modulation depths $|M_{GL}^{\omega}|$ and $|M_{PGL}^{\omega}|$ 
 as functions of the modulation frequency $\omega$ calculated using Eqs.~(21) with Eqs.~(19) and (20) for different carrier frequencies $\Omega/2\pi$ (different photon energies $\hbar\Omega$). It is assumed that $I = 2\times 10^{21}$~cm$^{-2}$s$^{-1}$ and, hence, $t_0 = 84$~ps, $t^{(1)} = 35$~ps,    $t^{(2)} = 21 $~ps , and    $t^{(3)} = 7$~ps (see Table II).                       
Other parameters are the same, as in Fig.~2, namely,  $\kappa = 4$, $a = 0.196$, and $\Delta = 400$~meV.

In the range of the photon energies, i.e., the carrier frequencies $\Omega < \omega_0$.
the modulation depth decreases with the modulation frequencies exceeding several GHz.
At $\Omega > \omega_0$, a marked decrease in $|M_{GL}^{\omega}|$
and $|M_{PGL}^{\omega}|$ takes place starting at more than 10 GHz.

It is interesting that at $\Omega > \omega_0$, the dependences of $|M_{GL}^{\omega}|$
and $|M_{PGL}^{\omega}|$ on the modulation frequency $\omega/2\pi$ exhibit maxima
at $\omega/2\pi$ about several MHz at which $|M_{GL}^{\omega}|$
and $|M_{PGL}^{\omega}|$ exceed $|M_{GL}^{0}|$
and $|M_{PGL}^{0}|$. This can be explained by a change in the phase shift  between the
quasi-Fermi and the effective temperature oscillations with a change of the modulation frequency.

\section{COMMENTS}

\subsection{Comparison of   GL and   PGL  photodetectors}

Using Eqs.~(8), (9), (13), (14), (19), and (20), we can arrive at the following ratios of the photocurrents normalized by the dark currents in the GL photoresistors
and the PGL photodiodes:  

\begin{eqnarray}\label{eq22}
\frac{\Delta J_{PGL}/J_{PGL}^{dark}}{\Delta J_{GL}/J_{GL}^{dark}} \sim
\frac{\delta J_{PGL}^{\omega}/J_{PGL}^{dark}}{\Delta J_{GL}^{\omega}/J_{GL}^{dark}}
\sim \biggl(\frac{\Delta_{GL}}{T_0}\biggr) \gg 1.
\end{eqnarray}

\subsection{Comparison of  GL and  PGL photodetectors  with  reverse-biased GLD photodiodes}

The photocurrents $\Delta J_{GL} = J_{GL} - J_{GL}^{dark}$
in the GL photodetectors under consideration and 
 $\Delta J_{GLD} =J_{GLD} - J_{GLD}^{dark}$   in the   p-i-n photodiodes~\cite{41}   at
a reverse strong bias providing the depletion of the GL with not too long  absorbing GL 
(with the spacing between the side contact  $2L \lesssim L_D$, where $L_D$ is the drift
length) can be estimated as follows~\cite{41}:

\begin{eqnarray}\label{eq23}
\frac{\Delta J_{GLD}}{2LH} = 2e \beta\, \tanh\biggl(\frac{\hbar\Omega - 2\mu}{4T}\biggr)\,I_{\Omega}.
\end{eqnarray}
The dark current associated with the carrier generation due to the interband absorption of optical phonons  in the GLDs is given by 

\begin{eqnarray}\label{eq24}
\frac{J_{GLD}^{dark}}{2LH} = 2eI.
\end{eqnarray}
Using Eq.~(12), for the ratios of the dc photocurrents (detector responsivities) normalized by the dark currents we obtain

\begin{eqnarray}\label{eq25}
\eta_{GL} = \frac{\Delta J_{GL}/J_{GL}^{dark}}{\Delta J_{GLD}/J_{GLD}^{dark}} 
\simeq \frac{1}{4a}\biggl(1 + a - \displaystyle\frac{\Omega}{\omega_0}\biggr),
\end{eqnarray}

\begin{eqnarray}\label{eq26}
\eta_{PGL} =\frac{\Delta J_{PGL}/J_{PGL}^{dark}}{\Delta J_{GLD}/J_{GLD}^{dark}} 
\simeq \frac{1}{2a}\biggl(\frac{\Delta}{2T_0}\biggr) 
 \,\biggl[1 +a - \frac{\Omega}{\omega_0}\nonumber\\
+ \biggl(\frac{T_0}{\hbar\omega_0}\biggr)\,\biggl(\frac{\Omega}{\omega_0} - 1 \biggr)\biggr].
\end{eqnarray}
One can see    from Eq.~(25) that
at $\Omega < \omega_0$, i.e., 
in the THz and far-IR spectral ranges   $\eta_{GL} \sim 1$. If $\Omega > \omega_0$ (mid- and near-IR ranges), $|\eta_{GL}|$ can markedly exceed unity.
Due to a large factor $(\Delta/2T_0)$, $|\eta_{PGL}|$
can be particularly large in a wide spectral range. Indeed, if,  for example,  $\Delta
= 400$~meV and
$\hbar\Omega = 40 - 100$~ meV, Eq.~(26) yields $\eta_{PGL} \simeq  9 - 18$.

%%%%%%%%%%%%%Responsivity
The GL-photoresistor and  PGL photodiode  responsivities $R_{GL} = \Delta J_{GL}/2LH I_{\Omega}$ and $R_{PGL} = \Delta J_{PGL}/2LH I_{\Omega}$
can be derived using Eqs.~(12) and (13):

\begin{eqnarray}\label{eq26}
R_{GL} \simeq \frac{\beta\,e}{16\pi\hbar\,a(2L)^2I}\biggl(\frac{eV\tau_0}{\hbar}\biggr)
\biggl(1 +a - \frac{\Omega}{\omega_0
}\biggr)\nonumber\\
\times\frac{\tanh(\hbar\Omega/4T_0)}{(\hbar\Omega/4T_0)},
\end{eqnarray}

\begin{eqnarray}\label{eq28}
R_{PGL} \simeq \frac{\beta\,eN}{4\pi\hbar\,a(2LH)I}
\biggl(\frac{eV}{T_0}\biggr) 
\biggl(\frac{\Delta}{2T_0}\biggr)
\exp\biggl(-\frac{\Delta}{2T_0}\biggr)\nonumber\\
\times\biggl[1 +a - \frac{\Omega}{\omega_0}
+ \biggl(\frac{T_0}{\hbar\omega_0}\biggr)\,\biggl(\frac{\Omega}{\omega_0} - 1 \biggr)\biggr]
\frac{\tanh(\hbar\Omega/4T_0)}{(\hbar\Omega/4T_0)}.
 \end{eqnarray}
Assuming $I = 2\times 10^{21}$~cm$^{-2}$s$^{-1}$, $\tau_0 = 1$~ps, $2L = 10~\mu$m, and $V = 100$~mV for  $\hbar\Omega = 100$~meV, we find $R_{GL} \simeq 0.04$~A/W. For  this photon energy, one obtains $R_{GLD} \simeq 0.17$~A/W.
Setting $2L = H = 10~\mu$m and $N = 100$, for the same voltage and photon energy,
we obtain $R_{PGL} \simeq 0.95\displaystyle \biggl(\frac{\Delta}{2T_0}\biggr)
\exp\biggl(-\frac{\Delta}{2T_0}\biggr)$~A/W. If  $\Delta = 125$~meV (corresponding to a smaller $\Delta/2T_0$ than in the above estimate of $\eta_{PGL}$), one obtains $R_{PGL} \simeq 0.19$~A/W. These estimates show that the  PGL photodiodes can surpass the GLD photodiodes in the parameter $\eta_{PGL}$ because of low values of the dark current, exhibiting, however, the responsivities of the same order of magnitude.

%Thus, the GL photoresistors and, in particular, the PGL photodiodes might surpass the %reverse biased GL photodiodes in the responsivity, but the former exhibit somewhat  %lower ultimate modulation frequencies (from a several to $10-
%15$ GHz) than the latter (tens of GHz~\cite{39}).

\subsection{Role of Auger processes}
Due to a smallness of the parameter $b$, the Auger processes weakly affect the characteristics of the photodetectors under consideration. This in particular, implies that despite a weak temperature dependence of the uniform GL conductivity, its photoresponse can be marked due to the deviation of the quasi-Fermi levels from the Dirac point cased by the 2DEHP heating or cooling. 
This  is contrast to the situation when a relatively dense  2DEHP is hot~\cite{42,43,44,45,46,47,48,49,50}, so that the Auger processes in GLs are rather effective~\cite{34}. The latter forces the quasi-Fermi levels be very close to the Dirac point ($\mu \simeq 0$)~\cite{51}.

\subsection{Assumptions} 

The screening lengths in GLs  with $\mu \simeq 0$ can be estimated as~\cite{51}

\begin{equation}\label{eq29}
l_{GL} = \frac{\kappa\hbar^2}{e^2}\frac{ v_W^2}{8\ln 2 T_0}.
\end{equation}
For $\kappa = 4$, from Eq.~(22) we obtain $l_{GL} \simeq 7.6$~nm 
This length is  much smaller than the characteristic carrier wave lengths $\lambda_{GL}=2\pi \hbar\,v_W/ T_0\simeq 150$~nm. 

This justifies the assumption of our model that the ionized impurities are effectively screened, so that the carrier scattering on such impurities, point defects, and acoustic phonons is a short-range scattering.

In Eq.~(2) we used the quantity $G$
 given by Eq.~(4). The latter ignores the contribution of the intraband photon absorption by the carriers (the Drude absorption). This is justified if (see, for example,~\cite{9})
 
 \begin{equation}\label{eq30}
 \tanh\biggl(\frac{\hbar\Omega}{4T_0}\biggr)  \gg \frac{D_0}{1 + 3\Omega^2\tau_0^2/\pi^2}, 
\end{equation}
where $D_0 =(4T_0\tau_0/\pi\hbar)$ stands for the Drude factor.
Inequality~(30) yields the following condition:

 \begin{equation}\label{eq31}
\hbar\Omega  \gg \sqrt{\frac{4\pi}{3}\frac{\hbar\,T_0}{\tau_0}} = \hbar\Omega_0. 
\end{equation}
For $T = 25$~meV and $\tau_0 = 0.1-1.0$~ps, one obtains $\hbar\Omega_0 \simeq 8 - 25$~meV. The photon energies assumed in the above calculations are much larger.
The generalization of our model for smaller $\hbar\Omega$ is rather simple.

The biased voltage can lead to some 2DEHP Joule heating. Considering that the Joule power in the GL per unit of its area $Q_J \simeq \sigma_0 E^2$, one can find the condition that the pertinent variation of the 2DEHP temperature 
$(T_J - T_0)/T_0 \ll 1$ if the electric field across the GL $E \ll E_J = 
\sqrt{\pi\,a\,I/\tau_0}(\hbar/e)(\hbar\omega_0/T_0)$. For the parameters used in the above estimates $E_J \simeq (175 - 555)$ ~V/cm.

Above we assumed that the operation of the PGL photodiode is determined by
the variations of the current across the nanoconstrictions.
Considering the voltage drop across the constriction area and the uniform GL region,
 the PGL photodiode net resistance $r_{PGL}$
 can be expressed as

 \begin{equation}\label{eq32}
r_{PGL} = \frac{2(L-l)}{H\sigma_{GL}} + r_{NC}.
 \end{equation}
Here $\sigma_{GL}$ is given by Eq.~(5) and 
$$
r_{NC} = \frac{\pi\hbar}{2e^2N}\exp\biggl(\frac{\mu + 
\Delta/2}{T}\biggr) 
$$ 
is the resistance of the perforated area for a parabolic potential distribution along the nanoconstrictions [see Eq.~(13) in~\cite{22}] with $N$ being the number of the nanoconstrictions. Hence, the variation of the nanoconstriction resistance (assumed above) is much larger than that of the uniform parts of the GL if

\begin{equation}\label{eq33}
\exp\biggl(\frac{\Delta}{2T_0}\biggr) \gg \frac{4NL}{H}\frac{\hbar}{T_0\tau_0},
\end{equation}
i.e.,

\begin{equation}\label{eq34}
\Delta >  2T_0 \ln\biggl[\frac{4N(L-l)}{H}\frac{\hbar}{T_0\tau_0}\biggr].
\end{equation}
Setting $2(L-l) = 10~\mu$m, $H = 10~\mu$m,  $N = 10$ ($h = H/N = 1~\mu$m) and $\tau_0 = 0.1$~ps or $N = 100$ and $\tau_0 = 1.0$~ps, for $T_0 = 25$~meV we find that, according to inequality~(33), a marked distinction in the GL and PGL photodetectors requires
$\Delta > 115$~meV.

\section{CONCLUSIONS}

We developed analytical models for the uncooled GL  and PGL photodetectors.  Their operation is linked  to
the correlated variations of the carrier density (i.e., quasi-Fermi energies) and effective temperature associated with the interband absorption of the incident modulated radiation. 
 We derived the modulation characteristic as a function of the carrier and modulation frequencies, demonstrated that the variation of both the carrier density and the effective temperature are essential for the
  GL- and PGL-based photodetectors, and showed that the value 2DEHP capacity 
 (markedly different form the classical value) affects the response time. 
The comparison of photodetectors under consideration 
showed that the presence of the energy barrier for the carriers in the nanoconstrictions in the PGL photodiodes promotes the higher photodetectors performance.
We also compared the  the GL- and PGL-based photodetectors with the interband photodiodes with the reverse-bias depleted GL. As shown the former can surpass 
the latter somewhat slower devices.

\section*{ACKNOWLEDGMENTS}

The Japan Society for Promotion of Science (KAKENHI $\#$16H06361), Japan;  RIEC Nation-Wide Collaborative Research Project $\#$H31/A01), Japan; 
the Russian  Foundation for Basic Research
(Grant $\#$18-29-02089), Russia; 
the Office of Naval Research, USA; the US Army Research Laboratory Cooperative Research Agreement, USA. 

\section*{DATA AVAILABILITY}
%\vspace{-5 mm}
The data that support the findings of this study are available from the corresponding author upon reasonable request.

\section*{Appendix A. The density and energy relaxation rates}
\setcounter{equation}{0}
\renewcommand{\theequation} {A\arabic{equation}}

The rates of the interband and intraband processes involving the optical phonons can be presented as:

\begin{equation}\label{eq1A}
R_{GL}^{inter}=I
\biggl[\exp\biggl(\frac{\hbar\omega_0}{T_0} - \frac{\hbar\omega_{0} -2\mu}{T}  \biggl) -1 \biggr], \\
\end{equation}

and 

\begin{equation}\label{eqA2}
R_{GL}^{intra}= a I\biggl[\exp\biggl(\frac{\hbar\omega_0}{T_0}-\frac{\hbar\omega_{0}}{T} \biggl)\biggr] -1.
\end{equation}
Here $I$ is the rate of the electron-hole pair generation due to the absorption of equilibrium  optical phonons in the GLs,
the quantities  $a$ is  the ratio of the rates of the intraband and interband transitions in the GLs  accompanied by the absorption of an optical phonon. This quantity is mainly determined by the  energy dependences of the densities of states around the Dirac point.
For GLs, at $|\mu|  \ll T$,    $a\simeq  (\pi\,T/\hbar\omega_0)^2 (1 + 2.19T/\hbar\omega_0)\simeq 0.196 $~\cite{4}.

According to the calculation~\cite{30}, at 
 $T_0 = 25$~meV, for the intra-valley and inter-valley optical phonons in GLs, $I \simeq (1 - 2)\times 10^{21}$~cm$^{-2}$s$^{-1}$~\cite{21}.
The characteristic time of the interband transitions with absorption of an optical phonon can be defined as $\tau_0 = \Sigma_0/I$, where $\Sigma_0 = (\pi/3){T_0\hbar\,v_W}$.

The rates of the Auger recombination-generation processes are assumed in the following form:

\begin{eqnarray}\label{eqA3}
R_{GL}^A = A_{GL}\biggl[\exp\biggl(\frac{2\mu}{T}\biggr) - 1\biggr],
\end{eqnarray}
where $A_{GL}$  is the rate of the electron-hole pair generation due to the impact Auger processes
in the 2DEHP with the equilibrium carrier density.
In GLs these processes are characterized by $t_A \propto A_{GL}^{-1}$ so that $b = t_0/t_A = A_{GL}/I$.
Since the time of the carrier recombination due to the spontaneous optical phonon emission $t_R \simeq t_0\exp(-\hbar\omega_0/T_0) \ll t_0$. The characteristic time $t_A$ can be estimated as $t_A \sim t_A^{R}
\exp(\hbar\omega_0/T_0)$. Setting $t_A^{R} \gtrsim (0.5 - 1.0) $~ps~\cite{34}, we find that the parameter $t_A (\simeq 1.5 -3)\times 10^3$~ps,
so that $b \simeq 0.028 - 0.056$, i.e., fairly small.

\section*{Appendix B. GL carrier and energy  densities versus carrier quasi-Fermi energy and effective temperature }
\setcounter{equation}{0}
\renewcommand{\theequation} {B\arabic{equation}}

The net carrier (electrons and holes) densities $\Sigma$  in the GL  in line with Eq.~(2) is given by

\begin{eqnarray}\label{eqB1}
 \Sigma = \frac{2}{\pi\hbar^2v_W^2}\int_0^{\infty}d\varepsilon \varepsilon
\biggl[\frac{1}{1 + \displaystyle\exp\biggl(\frac{\varepsilon - \mu_e}{T}\biggr)}\nonumber\\
 + \frac{1}{1 + \displaystyle\exp\biggl(\frac{\varepsilon - \mu_h}{T}\biggr)}\biggr].
\end{eqnarray}

 The density of the carrier energy in the 2DEHP can be calculated as

\begin{eqnarray}\label{eqB2}
{\cal E} =\frac{2}{\pi\hbar^2v_W^2}\int_0^{\infty}d\varepsilon\varepsilon^2
\biggl[\frac{1}{1 + \displaystyle\exp\biggl(\frac{\varepsilon - \mu_e}{T}\biggr)}\nonumber\\
 + \frac{1}{1 + \displaystyle\exp\biggl(\frac{\varepsilon - \mu_h}{T}\biggr)}\biggr].
\end{eqnarray}

In the undoped GLs, the electron and hole densities are equal to each other.
In this case, due to a symmetry of the valence and conduction bands, $\mu_e = \mu_h= \mu$. Except the situations when GLs and GBLs are strongly optically or injection pumped with the generation of cold or warm carriers,
 $|\mu_e + \mu_h| = 2|\mu|\ll T$. In such cases, Eqs.~(B1) and  (B2) yield~\cite{40}

\begin{eqnarray}\label{eqB3}
 \Sigma = \biggl(\frac{T}{\hbar\,v_W}\biggr)^2\biggl(\frac{\pi}{3} + \frac{4\ln 2}{\pi} \frac{\mu}{T}\biggr), 
 %\simeq \frac{\pi}{3}\biggl(\frac{T}{\hbar\,v_W}\biggr)^2,
\end{eqnarray}

\begin{eqnarray}\label{eqB4}
{\cal E}\simeq \frac{2T^3}{\pi\hbar^2v_W^2} \biggl[3\zeta(3) + \frac{\pi^2}{3}\frac{\mu}{T}\biggr],
%\simeq \frac{6\zeta(3)T^3}{\pi\hbar^2v_W^2},
\end{eqnarray}
where  $\zeta(x)$ is the Riemann zeta function: $\zeta(3) \simeq  3.61/2$.

Considering that the 2DEHP heat capacity in GLs is defined as
$C_{GL} = \partial {\cal E}_{GL}/\partial T$, from Eq.~(B4) we obtain

\begin{eqnarray}\label{eqB5}
C \simeq \frac{2T^2}{\pi\hbar^2v_W^2} \biggl[9\zeta(3) + \frac{2\pi^2}{3}\frac{\mu}{T}\biggr],
%\simeq \frac{6.57\pi}{3}\biggl(\frac{T}{\hbar\,v_W}\biggr)^2 
\end{eqnarray}
so that the 2DEHP capacity, $c = C/\Sigma $  (in units of the Boltzmann constant $k_B$),
  per one carrier is given by
\begin{eqnarray}\label{eqB8}
 c = \frac{54\zeta(3)}{\pi^2} \simeq 6.58.
\end{eqnarray}
This value is markedly different from the classical value for nondegenerate 2D systems $c_{2D} = 1$~\cite{40}.


\begin{thebibliography}{1}


%Heating/cooling. 

\bibitem{1} 
V. Ryzhii, M. Ryzhii, and T. Otsuji, \lq\lq Negative dynamic conductivity of graphene with optical pumping,\rq\rq
 J. Appl. Phys. {\bf 101}, 083114 (2007).


\bibitem{2} 
F. T. Vasko and V. Ryzhii, \lq\lq Photoconductivity of intrinsic
graphene,\rq\rq Phys. Rev. B {\bf 77}, 195433 (2008).


\bibitem{3} 
A. Satou, F. T. Vasko, and V. Ryzhii, \lq\lq Nonequilibrium carriers in intrinsic graphene under interband photoexcitation,\rq\rq
Phys. Rev. B {\bf 78} , 115431 (2008).


\bibitem{4} 
V. Ryzhii, M. Ryzhii,V. Mitin, A. Satou, and T. Otsuji, \lq\lq Effect of heating and cooling of photogenerated electron-hole
plasma in optically pumped graphene on population inversion,\rq\rq Jpn. J. Appl. Phys. {\bf 50}, 094001 (2011).


\bibitem{5}
  S. Boubanga-Tombet, S. Chan, T. Watanabe, A. Satou, V. Ryzhii, and T. Otsuji, \lq\lq
Ultrafast carrier dynamics and terahertz emission in optically pumped graphene at room temperature,\rq\rq
Phys. Rev. B {\bf 85},  035443 (2012). 



\bibitem{6}
  T. Li, L. Luo, M. Hupalo, J. Zhang, M. C. Tringides, J. Schmalian, and J.Wang, \lq\lq Femtosecond population inversion and stimulated emission of dense Dirac fermions in graphene,\rq\rq 
Phys. Rev. Lett. {\bf  108},  167401 (2012). 

\bibitem{7} 
 I. Gierz, J. C. Petersen, M. Mitrano, C. Cacho, I. E. Turcu, E. Springate, A. St{\"o}hr, A. K{\"o}hler, U. Starke, and A. Cavalleri, \lq\lq Snapshots of non-equilibrium Dirac carrier distributions in graphene,\rq\rq
 Nat. Mater. {\bf 12}, 1119 (2013).


\bibitem{8} 
J. N. Heyman, J. D. Stein, Z. S. Kaminski, A. R. Banman,
A. M. Massari, and J. T. Robinson, \lq\lq Carrier heating
and negative photoconductivity in graphene,\rq\rq 
J. Appl. Phys. {\bf 117},  015101 (2015).

 \bibitem{9}
 V. Ryzhii, D. S. Ponomarev, M. Ryzhii, V. Mitin, M. S. Shur, and T. Otsuji,
 \lq\lq  Negative and positive terahertz and infrared
photoconductivity in uncooled graphene,\rq\rq Opt. Mat. Express {\bf 9}, 585 - 597 (2019).


%BOLOMETERS

\bibitem{10} 
M. Shur, A.V. Muraviev, S. L. Rumyantsev, W. Knap,
G. Liu, and A. A. Balandin, \lq\lq Plasmonic and bolometric
terahertz graphene sensors,\rq\rq Proc. of 2013 IEEE Sensors
Conf., 978-1- 4673-4642-9/13/ 2013 IEEE pp. 1688-1690
(2013).








\bibitem{11}
J. Yan, M.-H. Kim,  J. A. Elle,  A.B. Sushkov,  G.S. Jenkins,  H.M. Milchberg, 3, M.S. Fuhrer, and H.D. Drew, \lq\lq
Dual-gated bilayer graphene hot electron bolometer,\rq\rq 
Nat. Nanotech. {\bf 7}, 472--478 (2012).

\bibitem{12}
V. Ryzhii, T. Otsuji, M. Ryzhii, N. Ryabova, S. O. Yurchenko, V. Mitin, and M. S. Shur, \lq\lq Graphene terahertz
uncooled bolometers,\rq\rq J. Phys. D: Appl. Phys. {\bf 46}, 065102 (2013).


\bibitem{13}
Q Han, T. Gao, R. Zhang, Y. Chen, J.Chen, G. Liu, Y. Zhang, Z. Liu, X. Wu,
and   D.  Yu,
\lq\lq  Highly sensitive hot electron bolometer based on disordered graphene,\rq\rq
Sci. Reports {\bf 3}, 3533 (2013).


\bibitem{14}
X. Cai, A. B. Sushkov, R J. Suess, M. M. Jadidi, G. S. Jenkins, L. O. Nyakiti, R. L. Myers-Ward, S. Li, J. Yan, D. K. Gaskill, T. E. Murphy, H. D. Drew, and  M. S. Fuhrer,
\lq\lq Sensitive room-temperature terahertz detection via the photothermoelectric effect in graphene,\rq\rq
Nat. Nanotech. {\bf 9}, 814-819 (2014). 

\bibitem{15} 
X. Du, D. E. Prober, H. Vora, and Ch. B. Mckitterick,
\lq\lq Graphene-based bolometers,\rq\rq
Graphene 2D Mater. {\bf 1}, 1-22 (2014).


\bibitem{16}
A. El Fatimy, R. L. Myers-Ward, A. K. Boyd, K. M. Daniels, D. K. Gaskill, and P. Barbara,
\lq\lq Epitaxial graphene quantum dots for high-performance terahertz bolometers,\rq\rq
Nat. Nanotechnol. {\bf 11}, 335-338 (2016).



\bibitem{17}
S. Yuan, R. Yu, C. Ma, B. Deng, Q. Guo, X. Chen, C. Li, C. Chen, K. Watanabe, T. Taniguchi, F. J. García de Abajo, and F. Xia,
\lq\lq Room temperature graphene mid-infrared bolometer with a broad operational wavelength range,\rq\rq
ACS Photonics {\bf 7}, 1206-1215 (2020).

\bibitem{18}
V. Ryzhii, M. Ryzhii, D. Ponomarev, V. G. Leiman, V.
Mitin, M. Shur, and T. Otsuji, 
\lq\lq Negative photoconductivity
and hot-carrier bolometric detection of terahertz
radiation in graphene-phosphorene hybrid structures,\rq\rq
J. Appl. Phys. 125, 151608 (2019).


\bibitem{19} 
A. Blaikie, D. Miller, and   B. J. Alem{\'a}n,
\lq\lq A fast and sensitive room-temperature graphene nanomechanical bolometer,\rq\rq
Nat. Comm. {\bf 10}, 4726 (2019). 
\bibitem{20} 
G.-H. Lee, D. K. Efetov, L. Ranzani, E. Walsh, J. Crossno, T. A. Ohki, T. Taniguchi, K. Watanabe, P. Kim, D. Englund, and K. C. Fong, \lq\lq Graphene-based Josephson junction microwave bolometer,\rq\rq
Nature {\bf 586}, 42 – 46 (2020).




\bibitem{21} 
G. S. Simin, M. Islam, M. Gaevski, J. Deng, R. Gaska, and M. S. Shur, \lq\lq Low RC-constant perforated-channel HFET,\rq\rq IEEE Electron Device Lett. {\bf 35}, 452-454 (2014). 
 
\bibitem{22}
V. Ryzhii, M. Ryzhii, M. S. Shur, V. Mitin, A. Satou, and T. Otsuji,
\lq\lq Resonant plasmonic terahertz detection in graphene split-gate field-effect transistors with lateral p-n junctions,
\rq\rq J. Phys. D: Appl. Phys. {\bf 49}, 315103 (2016). 

\bibitem{23}
R. P. Panmand, P.Patil, Y Sethi, S. R. Kadam, M. V. Kulkarni,S. W. Gosavi, N. R. Munirathran,
and B. B. Kale,
\lq\lq Unique perforated graphene derived fromBougainvilleaflowers for high-powersupercapacitors: a green approach,\rq\rq
Nanoscale {\bf 9}, 4801 (2017).



\bibitem{24}
A. Guirguisa, J. W. Maina, L. Kong, L. C. Henderson, A. Rana, L. H. Li, 
M. Majumder, L. F. Dumee,
\lq\lq Perforation routes towards practical nano-porous graphene andanalogous materials engineering
\rq\rq Carbon {\bf 155}, 660 (2019).




\bibitem{25}
R. A. Suris and V. A. Fedirko, \lq\lq Heating photoconductivity in a semiconductor with a superlattice,\rq\rq
Sov. Phys. Semicond. {\bf 12}, 629 (1978).

\bibitem{26} E. A. Shaner, A. D. Grine, M. C. Wanke, M. Lee, J. R. Reno, and S. J. Allen, \lq\lq
Far-infrared spectrum analysis using plasmon modes in a quantum-well transistor,
\rq\rq IEEE Photonics Technol. Lett. {\bf 18}, 1925 (2006).


\bibitem{27}  
V. Ryzhii, A. Satou, T. Otsuji, and M. S. Shur, \lq\lq
Plasma mechanism of resonant terahertz detection in a two-dimensional electron channel with split gates,\rq\rq
J. Appl. Phys. {\bf 103}, 014504 (2008).

\bibitem{28}
V. Ryzhii, A. Satou, T. Otsuji, M. Ryzhii, V. Mitin, and M. S. Shur,
\lq\lq Graphene vertical hot-electron terahertz detectors,\rq\rq
J. Appl. Phys. {\bf 116}, 114504 (2014).
 

\bibitem{29}
M. Ryzhii, V. Ryzhii, V. Mitin, M. S. Shur, and T. Otsuji,
\lq\lq Vertical hot-electron terahertz  detectors based on black-As$_{1-x}$P$_x$/graphene/black-As$_{1-x}$P$_x$ heterostructures,\rq\rq
Sensors and Materials {\bf 31}, 2271-2279  (2019).








  
\bibitem{30} 
 F. Rana, P. A. George , J. H. Strait, S. Sharavaraman, M. Charasheyhar, and M. G. Spencer, \lq\lq Carrier recombination and generation rates for intravalley and intervalley phonon scattering in graphene,\rq\rq
 Phys. Rev. B {\bf 79}, 115447 (2009).

\bibitem{31}
H. Wang, J. H. Strait, P. A. George, S. Shivaraman, V. D. Shields, M. Chandrashekhar, J. Hwang, F. Rana, M. G. Spencer, C. S. Ruiz-Vargas, and J. Park, \lq\lq Ultrafast relaxation dynamics of hot optical phonons in graphene,\rq\rq
 Appl. Phys. Lett. {\bf  96}, 081917 (2010). 
 
 \bibitem {32}
J. M. Iglesias,  M. J. Martín, E. Pascual,  and R. Rengel, \lq\lq Hot carrier and hot phonon coupling during ultrafast relaxation of photoexcited electrons in graphene,\rq\rq
Appl. Phys. Lett. {\bf 108}, 043105 (2016). 
%https://doi.org/10.1063/1.4940902, Google ScholarScitation

 
%Auger

\bibitem{33}
M. S. Foster and I. L. Aleiner, \lq\lq Slow imbalance relaxation and thermoelectric transport 
in graphene,\rq\rq 
Phys. Rev. B {\bf 79}, 085415 (2009). 

\bibitem{34} 
G. Alymov, V. Vyurkov, V. Ryzhii, A. Satou, and D. Svintsov, \lq\lq 
Auger recombination in Dirac materials: A tangle of many-body effects,\rq\rq
Phys. Rev. B {\bf  97}, 205411 (2018). 

\bibitem{35}
V. Ryzhii, T. Otsuji, M. Ryzhii,  V. E. Karasik, and M. S. Shur,
\lq\lq  Negative terahertz conductivity and amplification of surface plasmons in graphene–black phosphorus injection laser heterostructures,\rq\rq
 Phys. Rev. B {\bf 100} 115436 (2019).

\bibitem{36}
A. Shik, {\it Quantum Wells: Physics and Electronics of Two-Dimensional systems} (World Scientific Singapore, 1997)

\bibitem{37}
V. V. Mitin, D. I. Sementsov, and N. Vagidov, {\it Quantum Mechanics for Nanostructures} (Cambridge Univ. Press, Cambridge, 2010).

\bibitem{38}
G. Liang, N. Neophytou, M. S. Lundstrom,
and D E. Nikonov, \lq\lq
Contact effects in graphene nanoribbon
transistors,\rq\rq Nano Lett. {\bf 8}, 1819-1824
(2008).




\bibitem{39}
K.-T. Lam and G. Liang,
\lq\lq Electronic structure of bilayer graphene nanoribbon and its device application: A computational study,\rq\rq NanoSci. and Technol. {\bf 57}, 509-527 (2012).
%DOI: 10.1007/978-3-642-22984-8-16

\bibitem{40}
V. Ryzhii, M. Ryzhii, T. Otsuji, V. Mitin, and M. S. Shur,
\lq\lq Heat capacity of nonequilibrium electron-hole plasma in graphene layers and
graphene bilayers,\rq\rq
arXiv:2011.03739 [cond-mat.mes-hall].


\bibitem{41}  
V. Ryzhii, M. Ryzhii, V. Mitin, and T. Otsuji, \lq\lq
Terahertz and infrared photodetection using p-i-n multiple-graphene layer structures,\rq\rq
J. Appl. Phys. {\bf 107}, 054512 (2010). 



\bibitem{42}
M. Freitag, H.-Y. Chiu, M. Steiner, V. Perebeinos, and P. Avouris,
"Thermal infrared emission from biased graphene,"
Nat. Nanotech. {\bf 5}, 497--501 (2010).


\bibitem{43}
Y. D. Kim, H. Kim, Y. Cho, Ji H. Ryoo, C.-H. Park, P. Kim, Y. S. Kim, S. Lee, Y. Li, S.-N. Park, Y. S. Yoo,  D. Yoon, V. E. Dorgan, E. Pop, T. F. Heinz, J. Hone, S.-H. Chun, H.  Cheong, S. W. Lee, M.-Ho Bae, and Y. D. Park,   
\lq\lq Bright visible light emission from graphene,\rq\rq
Nat. Nanotech. {\bf 10}, 676--681 (2015).

\bibitem{44}
H. R.Barnard, E. Zossimova, N. H. Mahlmeister, L. M. Lawton, I. J. Luxmoore, and G. R. Nash,
\lq\lq Boron nitride encapsulated graphene infrared emitters,\rq\rq
Appl. Phys. Lett. {\bf 108}, 131110 (2016).
% https://doi.org/10.1063/1.4945371 
%%% T 440 - 530 K, P ~(5- 20)kW/cm$^2$.

\bibitem{45}
S.-K. Son, M. \u{S}\,i\,\u{s}\,kins, C. Mullan, J. Yin, V. G. Kravets, A. Kozikov, S. Ozdemir, M. Alhazmi, M. Holwill, K. Watanabe, T. Taniguchi, D. Ghazaryan, K. S. Novoselov, V. I. Fal’ko, and A. Mishchenko,
\lq\lq Graphene hot-electron light bulb: incandescence from hBN encapsulated graphene in air,
\rq\rq
2D Materials {\bf 5}, 011006 (2017).

\bibitem{46}
H. M. Dong, W. Xu, and  F. M. Peeters,
\lq\lq Electrical generation of terahertz blackbody radiation from graphene,\rq\rq
 Opt. Express {\bf 26}(19), 24621--24626 (2018).



\bibitem{47}
R.-J. Shiue, Y. Gao, C. Tan, C. Peng, J. Zheng, D. K. Efetov,  Y. D. Kim, J.  Hone, and D. Englund,
\lq\lq Thermal radiation control from hot graphene electrons coupled to a photonic crystal nanocavity,\rq\rq
Nat. Commun. {\bf 10}, 109 (2019).


\bibitem{48}
F. Luo, Y. Fan, G. Peng, S. Xu, Y. Yang, K. Yuan, J. Liu, W. Ma, W. Xu, Z. H.  Zhu, X.-Ao Zhango, A. Mishchenko, Yu Ye, H. Huang, Z. Han, W. Ren, K. S. Novoselov, M. Zhu, and S. Qin,
\lq\lq Graphene thermal emitter with enhanced Joule heating and localized light emission in air,\rq\rq
ACS Photonics {\bf 6}, 2117--2125 (2019).


\bibitem{49} 
Y. D. Kim, Y. Gao, R.-J. Shiue, L. Wang, O. B. Aslan, M.-Ho Bae, H. Kim, D. Seo, H.-J. Choi, S. H. Kim, A. Nemilentsau, T. Low, C. Tan, D. K. Efetov, T. Taniguchi, K. Watanabe, K. L. Shepard, T. F. Heinz, D. Englund, and J. Hone,
\lq\lq Ultrafast graphene light emitters,\rq\rq
Nano Lett. {\bf 18}, 934--940 (2018).

\bibitem{50}
V. Ryzhii, T. Otsuji, M. Ryzhii, V. Leiman, P. P. Maltsev, V. E. Karasik, V. Mitin, and M. S. Shur, \lq\lq Theoretical analysis of injection driven thermal light emitters based on graphene encapsulated by hexagonal boron nitride,\rq\rq
Opt. Mat. Express {\bf 11}, 468 (2021). 

\bibitem{51}
V. Ryzhii, M. Ryzhii, D. Svintsov, V. Leiman, P. P. Maltsev,  D. S. Ponomarev,
V. Mitin, M. S. Shur, and T. Otsuji, \lq\lq
Real-space-transfer mechanism of negative differential conductivity in gated
graphene-phosphorene hybrid structures: Phenomenological heating model,\rq\rq
J. Appl. Phys. {\bf 124}, 114501 (2018).


\end{thebibliography}
\end{document}